\documentclass[]{aastex631}
\usepackage{amsmath}
\usepackage{comment}
\usepackage{textgreek}
\usepackage{multirow}

\shorttitle{}
\shortauthors{Amenomori et al., 2022}
\graphicspath{{./}{figures/}}

\begin{document}

\title{Measurement of the Gamma-Ray Energy Spectrum beyond 100 TeV from the HESS J1843-033 Region}

\author{M. Amenomori}
\affiliation{Department of Physics, Hirosaki University, Hirosaki 036-8561, Japan}
\author{S. Asano}
\affiliation{Department of Physics, Shinshu University, Matsumoto 390-8621, Japan}
\author{Y. W. Bao}
\affiliation{School of Astronomy and Space Science, Nanjing University, Nanjing 210093, China}
\author{X. J. Bi}
\affiliation{Key Laboratory of Particle Astrophysics, Institute of High Energy Physics, Chinese Academy of Sciences, Beijing 100049, China; hukongyi@ihep.ac.cn, huangjing@ihep.ac.cn}
\author{D. Chen}
\affiliation{National Astronomical Observatories, Chinese Academy of Sciences, Beijing 100012, China}
\author{T. L. Chen}
\affiliation{Department of Mathematics and Physics, Tibet University, Lhasa 850000, China}
\author{W. Y. Chen}
\affiliation{Key Laboratory of Particle Astrophysics, Institute of High Energy Physics, Chinese Academy of Sciences, Beijing 100049, China; hukongyi@ihep.ac.cn, huangjing@ihep.ac.cn}
\author{Xu Chen}
\affiliation{Key Laboratory of Particle Astrophysics, Institute of High Energy Physics, Chinese Academy of Sciences, Beijing 100049, China; hukongyi@ihep.ac.cn, huangjing@ihep.ac.cn}
\affiliation{National Astronomical Observatories, Chinese Academy of Sciences, Beijing 100012, China}
\author{Y. Chen}
\affiliation{School of Astronomy and Space Science, Nanjing University, Nanjing 210093, China}
\author{Cirennima}
\affiliation{Department of Mathematics and Physics, Tibet University, Lhasa 850000, China}
% ...
%\email{Corresponding authors' email addresses: katosei@icrr.u-tokyo.ac.jp, hukongyi@ihep.ac.cn, takita@icrr.u-tokyo.ac.jp, huangjing@ihep.ac.cn}

\begin{abstract}
  HESS J1843-033 is a very-high-energy gamma-ray source whose origin remains unidentified. This work presents, for the first time, the energy spectrum of gamma rays beyond $100\, {\rm TeV}$ from the HESS J1843-033 region using the data recorded by the Tibet air shower array and its underground muon detector array. A gamma-ray source with an extension of $0{\fdg}34 \pm 0{\fdg}12$ is successfully detected above $25\, {\rm TeV}$ at $(\alpha,\, \delta) = (281{\fdg}09\pm 0{\fdg}10,\, -3{\fdg}76\pm 0{\fdg}09)$ near HESS J1843-033 with a statistical significance of $6.2\, \sigma$, and the source is named TASG J1844-038. The position of TASG J1844-038 is consistent with those of HESS J1843-033, eHWC J1842-035, and LHAASO J1843-0338. The measured gamma-ray energy spectrum in $25\, {\rm TeV} < E < 130\, {\rm TeV}$ is described with ${\rm d}N/{\rm d}E = (9.70\pm 1.89)\times 10^{-16} (E/40\, {\rm TeV})^{-3.26\pm 0.30}\, {\rm TeV}^{-1} {\rm cm}^{-2} {\rm s}^{-1}$, and the spectral fit to the combined spectra of HESS J1843-033, LHAASO J1843-0338, and TASG J1844-038 implies the existence of a cutoff at $49.5\pm 9.0\, {\rm TeV}$. Associations of TASG J1844-038 with SNR G28.6-0.1 and PSR J1844-0346 are also discussed in detail for the first time. %Pion-decay gamma rays generated by hadronic interactions between molecular clouds and cosmic rays accelerated by SNR G28.6-0.1 can explain the observed extension of TASG J1844-038. On the other hand, if associated with PSR J1844-0346, the gamma-ray emission of TASG J1844-038 will be generated by Inverse Compton Scattering off the cosmic microwave background photons by high-energy electrons.
\end{abstract}


\section{Introduction} \label{sec:intro}
The cosmic-ray energy spectrum has the so-called knee at about $4\, {\rm PeV}$ \citep{Kristiansen_1958, Hoerandel_2004, Amenomori_et_al_2008}, and its origin is still unknown. Searches for a PeVatron, an accelerator of such PeV cosmic rays in the Galaxy, are performed through observations of sub-PeV gamma rays ($E>100\, {\rm TeV}$) stemming from decays of neutral pions generated by hadronic interactions of cosmic rays with the interstellar medium. So far about a dozen of sources have been discovered in the sub-PeV energy range (for example, see \citet{TibetCrab, HAWC56TeV, LHAASO100TeV}), and one of these sources, a supernova remnant (SNR) G106.3+2.7, is a candidate for a PeVatron because of good spatial correlation between the distributions of gamma rays and molecular clouds \citep{HAWCSNRG106, TibetSNRG106}. However, the maximum energy of cosmic-ray protons accelerated by SNR G106.3+2.7 is estimated at $\simeq 0.5\, {\rm PeV}$ lower than the knee energy range, and the robust detection of a PeVatron remains inconclusive. In many cases, gamma rays of leptonic origin also significantly contribute to the total gamma-ray flux of sources, making the situation confusing. On the other hand, the observation of sub-PeV Galactic diffuse gamma rays provides evidence that PeVatrons exist (and/or existed) in the Galaxy \citep{TibetDiffuse}. The result is consistent with the theory presented by \citet{Lipari_and_Vernetto_PRD_2018} arguing that gamma rays of hadronic origin dominate the total Galactic diffuse gamma-ray flux in the sub-PeV range. These facts motivate us to directly specify PeVatrons in the Galaxy.

HESS J1843-033 is one of the unidentified gamma-ray sources discovered by the H.E.S.S. Galactic plane survey \citep{HGPS2008, HGPS2018}. The energy spectrum was measured up to $30\, {\rm TeV}$ and described with a power-law function with an index of $\simeq -2.2$. HAWC also reported the discovery of 2HWC J1844-032 and eHWC J1842-035 near HESS J1843-033 \citep{2HWCCatalog, HAWC56TeV}, and for the latter, they estimated the integral gamma-ray flux above $56\, {\rm TeV}$. Moreover, LHAASO discovered LHAASO J1843-0338 and provided a flux point at $100\, {\rm TeV}$ \citep{LHAASO100TeV}. Theoretically, the nature of the gamma-ray emission has been discussed by \citet{Sudoh_et_al_2021} and \citet{Huang_et_al_2022}, but their claims are still inconclusive because the gamma-ray energy spectrum is not systematically measured above the several tens of TeV range and its characteristics are thus not revealed yet. This paper presents the first observational result of the energy spectrum of gamma rays beyond $100\, {\rm TeV}$ from the HESS J1843-033 region and discusses the origin of the gamma-ray emission considering its possible associations with nearby celestial objects.

\section{Experiment and data analysis} \label{sec:experiment}
The Tibet air shower array (AS array) has been operating since 1990 in Yangbajing ($90{\fdg}522\, {\rm E}$, $30{\fdg}102 {\rm N}$, $4,300\, {\rm m}$ a.s.l.) in Tibet, China \citep{Amenomori_et_al_1992, Amenomori_et_al_1999, Amenomori_et_al_2009} and currently consists of a surface AS array and an underground muon detector array (MD array, \citet{TibetCrab}). The AS array comprises 597 plastic scintillation detectors, each with a detection area of $0.5\, {\rm m}^2$, and covers an area of $65,700\, {\rm m}^2$. The MD array with a total area of $3,400\, {\rm m}^2$ consists of water-Cherenkov type detectors to collect Cherenkov light emitted by penetrating shower muons in the water layer. Each detector of the MD array is made of concrete and is located 2.4 m underground. The soil overburden and the concrete ceiling of the MD detectors correspond to the total thickness of $\simeq 550\, {\rm g}\, {\rm cm}^{-2}$, allowing only shower muons with $E\gtrsim 1\, {\rm GeV}$ to reach the water layer of the array for vertical incidence. Using the MD array thus enables us to improve the experiment's sensitivity to sub-PeV gamma rays by more than one order of magnitude by rejecting more than $99.9\%$ of hadronic cosmic rays while keeping $90\%$ of gamma rays in the sub-PeV energy range \citep{Sako_et_al_2009, TibetCrab}. A trigger is issued when any four plastic scintillation detectors detect more than $0.6$ particles within a time window of $600\, {\rm ns}$.

In this work, data taken from 2014 February to 2017 May (719 live days) are analyzed. Reconstruction methods for recorded AS events and selection criteria imposed on the events are the same as \citet{TibetCrab} except for the following two points. First, the maximum zenith angle of the incoming direction of analyzed events is changed from $40^{\circ}$ to $50^{\circ}$ because the meridian zenith angle of HESS J1843-033 is $\simeq 33^{\circ}$ at the site of the experiment. This extension helps us increase the exposure to the HESS J1843-033 region by $\simeq 70\%$ and improve statistics of gamma rays from that region. Second, the muon-cut condition optimized for this work is used (see Appendix \ref{app:MC}). After the selection, events are analyzed, binned in energy with five logarithmically equal bins per decade. For the Monte Carlo simulation performed in this work, see Appendix \ref{app:MC}.

\section{Results} \label{sec:res}
\begin{figure}[h]
  \centering
  \includegraphics[scale=0.6]{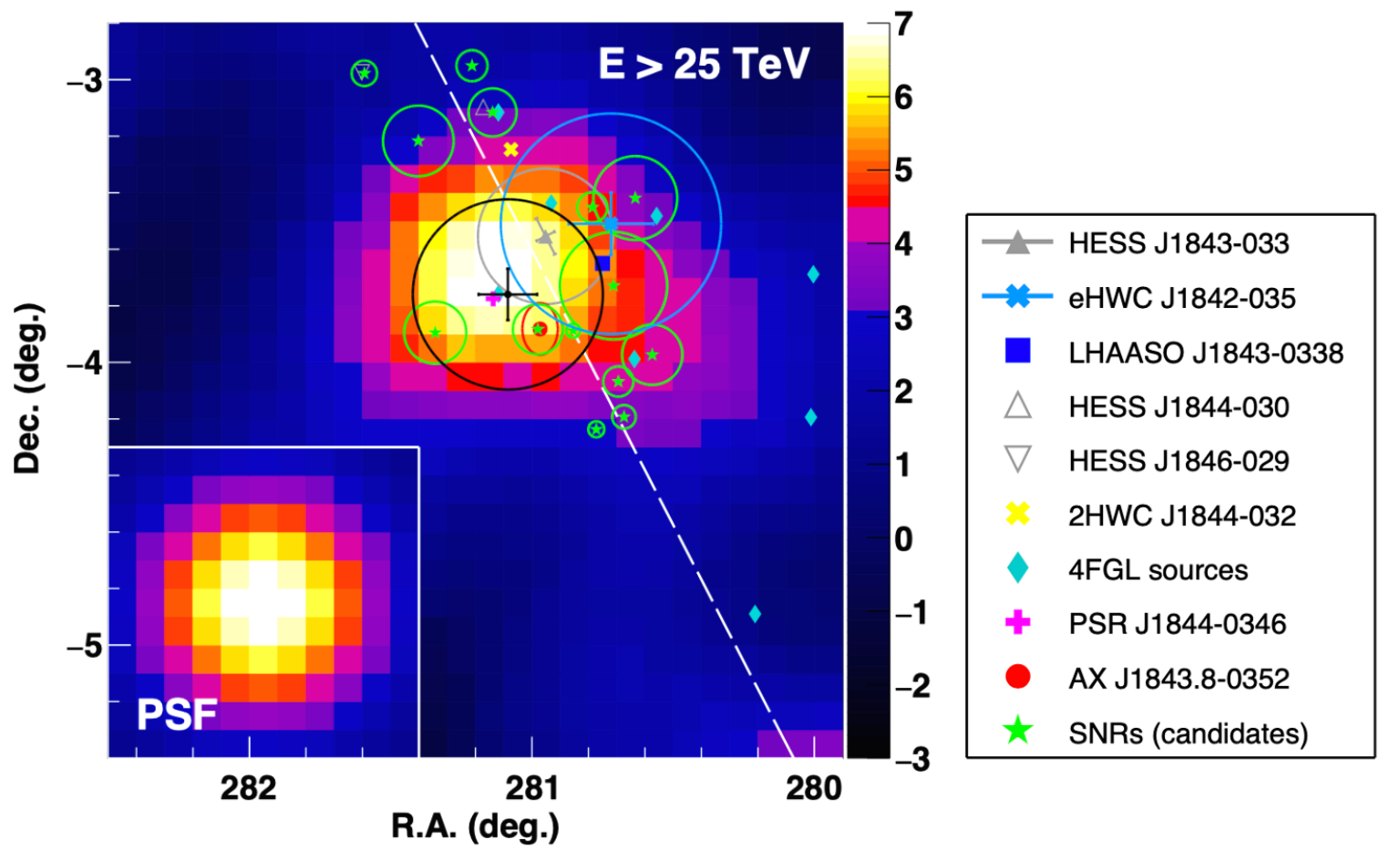}
    \caption{Significance map of TASG J1844-038 above $25\, {\rm TeV}$ smoothed with the point-spread function (PSF). The black cross shows the source center and its statistical errors in the right ascension and declination, and the source extension ($\sigma_{\rm ext}$, see the text) is denoted by the black circle. The Galactic plane is drawn by the white dashed line. The lower-left inset denotes the PSF above $25\, {\rm TeV}$. Positions and extensions of nearby sources listed in the right legend are indicated in the same way as for TASG J1844-038. For HESS J1843-033 and eHWC J1842-035, the statistical errors on their positions are also shown. Data of H.E.S.S. sources are taken from \citet{HGPS2018}, HAWC sources \citet{2HWCCatalog, HAWC56TeV}, LHAASO J1843-0338 \citet{LHAASO100TeV}, 4FGL sources \citet{4FGLCatalog}, PSR J1844-0346 \citet{C_J_Clerk_et_al_2017}, AX J1843.8-0352 \citet{Bamba_et_al_2001}, and SNRs (and candidates) \citet{Anderson_et_al_2017}.}
    \label{sigmap1}
\end{figure}

A gamma-ray source is successfully detected above $25\, {\rm TeV}$ near HESS J1843-033 with a statistical significance of $6.2\, \sigma$ calculated from Equation (17) of \citet{LiMa1983}. Figure \ref{sigmap1} shows the significance map of the observed gamma rays. A two-dimensional maximum likelihood analysis is performed to estimate the source center assuming an axisymmetric Gaussian distribution for the distribution of the gamma rays. The resultant center is $(\alpha,\, \delta) = (281{\fdg}09\pm 0{\fdg}10,\, -3{\fdg}76\pm 0{\fdg}09)$ in the J2000 equatorial coordinates. The source is named TASG J1844-038, and hereafter this name is used accordingly. Table \ref{tab:posinfo} presents the positions of TASG J1844-038 and nearby gamma-ray sources and shows that the position of TASG J1844-038 is statistically consistent with those of HESS J1843-033, eHWC J1842-035, and LHAASO J1843-0338. On the other hand, TASG J1844-038 deviates from HESS J1844-030 and HESS J1846-029 at the $3.2\, \sigma$ and $4.5\, \sigma$ levels, respectively, making its associations with these sources unlikely.
\begin{deluxetable*}{cchccccc}
\tablenum{1}
\tablecaption{Positions and extensions of TASG J1844-038 and nearby gamma-ray sources. {\it $\alpha$} and {\it $\delta$} are the right ascension and declination, respectively, in the J2000 equatorial coordinates. {\it $R_{0.68}$} denotes the error radius of a source position at the $68\%$ confidence level (see Appendix \ref{app:calcdev}). For the source extension of TASG J1844-038, see the text. Numbers enclosed in the parentheses in the {\it Angular distance to TASG J1844-038} column denote the significance of a positional deviation between TASG J1844-038 and a source evaluated with their $R_{0.68}$'s. Data of the nearby sources are taken from \citet{HGPS2018}, \citet{HAWC56TeV}, and \citet{LHAASO100TeV}. \label{tab:posinfo}}
\tablehead{
  \colhead{Source name} & \colhead{${\alpha} (^{\circ})$} & \nocolhead{name} & \colhead{$\delta (^{\circ})$}  & \colhead{$R_{0.68}$ ($^{\circ}$)} & \colhead{Extension ($^{\circ}$)} & \colhead{Angular distance to}\\
  \colhead{} & \colhead{} & \nocolhead{} & \colhead{} & \colhead{} & \colhead{} & \colhead{TASG J1844-038 ($^{\circ}$)}\\
}
\startdata
%TASG J1844-038 & $281.09$ & name & $-3.76$ &  $0.21$ & $0.35 \pm 0.11$ & -\\
TASG J1844-038 & $281.09$ & name & $-3.76$ &  $0.21$ & $0.34 \pm 0.12$ & -\\
HESS J1843-033 & $280.95$ & name & $-3.55$ & 0.12 & $0.24 \pm 0.06$ & 0.25 ($1.0\, \sigma$)\\
HESS J1844-030 & $281.17$ & name & $-3.10$ & 0.023 & $0.02\pm 0.013$ & 0.67 ($3.2\, \sigma$)\\
HESS J1846-029 & $281.60$ & name & $-2.97$ & 0.015 & $0.01\pm 0.013$ & 0.94 ($4.5\, \sigma$)\\
eHWC J1842-035 & $280.72$ & name & $-3.51$ & 0.30 & $0.39 \pm 0.09$ & 0.44 ($1.2\, \sigma$)\\
LHAASO J1843-0338 & $280.75$ & name & $-3.65$ & $0.16 $ & -$^{*}$ & 0.35 ($1.4\, \sigma$)\\
\enddata
\tablecomments{* The source extension is not published.} % [4] Pointing systematics of H.E.S.S., HAWC, and LHAASO, and this work are taken from \citet{HGPS2018}, \citet{2HWCCatalog}, \citet{LHAASOCrab}, and \citet{MoonShadow}, respectively.}
\end{deluxetable*}

\begin{figure}
  %\plotone{phisq.pdf}
  \plotone{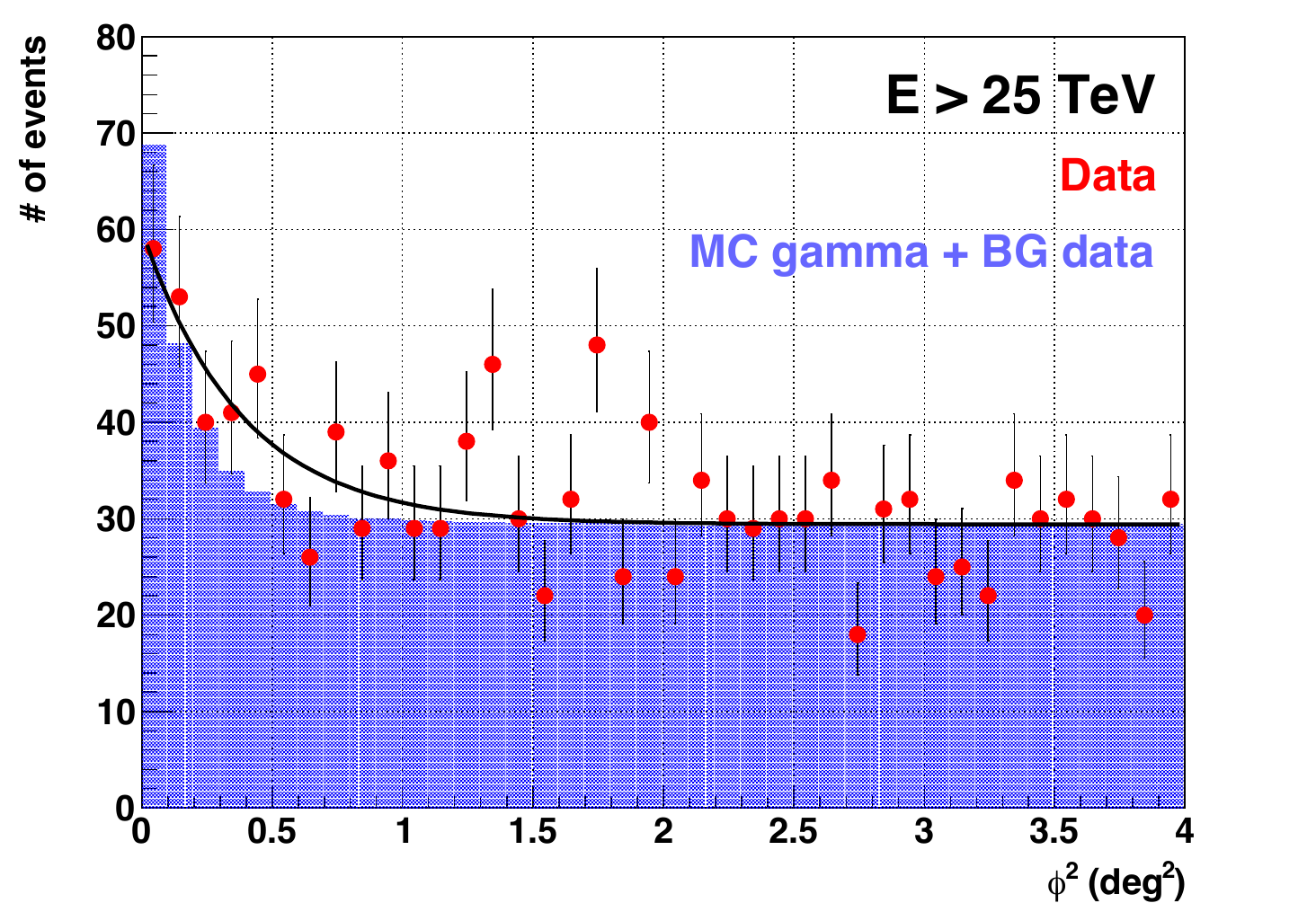}
  \caption{Distribution of events above $25\, {\rm TeV}$ around the center of TASG J1844-038. The horizontal axis $\phi^2$ denotes the square of the angle between the center of TASG J1844-038 and the incoming direction of events. The blue histograms are constructed from background events in OFF regions plus Monte Carlo gamma-ray events from a point source normalized to the number of excess counts in the ON-source region. The solid black curve shows the best-fit function to the data formulated as Equation \eqref{eq:phi2}.}
  \label{phisq}
\end{figure}

Figure \ref{phisq} presents the distribution of events above $25\, {\rm TeV}$ around the center of TASG J1844-038. The horizontal axis $\phi^2$ denotes the square of the angle between the center of TASG J1844-038 and the incoming direction of events. The blue histograms are constructed from background events in OFF regions plus Monte Carlo gamma-ray events from a point source normalized to the number of excess counts in the ON-source region. The source extension is estimated by fitting the following Gaussian function to the observed number of events:
\begin{equation}\label{eq:phi2}
  G({\phi}^2; A, \sigma_{\rm ext}) = A\, {\rm exp}\bigg(-\frac{\phi^2}{2(\sigma^2_{\rm ext}+\sigma^2_{\rm psf})}\bigg) + N_{\rm bg}
\end{equation}
where $A$ is a normalization constant, $\sigma_{\rm ext}$ the extension of TASG J1844-038, $\sigma_{\rm psf} = 0{\fdg}28$ the radius of the point-spread function for gamma rays above $25\, {\rm TeV}$ that follow a power-law energy spectrum with an index of $3.0$, and $N_{\rm bg} = 29.4$ the number of background events. The best-fit result is shown by the black curve in Figure \ref{phisq}, leading to $\sigma_{\rm ext} = 0{\fdg}34 \pm 0{\fdg}12$ with a reduced chi-square of $\chi^2/{\rm d.o.f.} = 39.5\, /\, 38$.

\begin{figure}
  %\plotone{enesp_0.35deg_2.pdf}
  %\plotone{enesp.pdf}
  \plotone{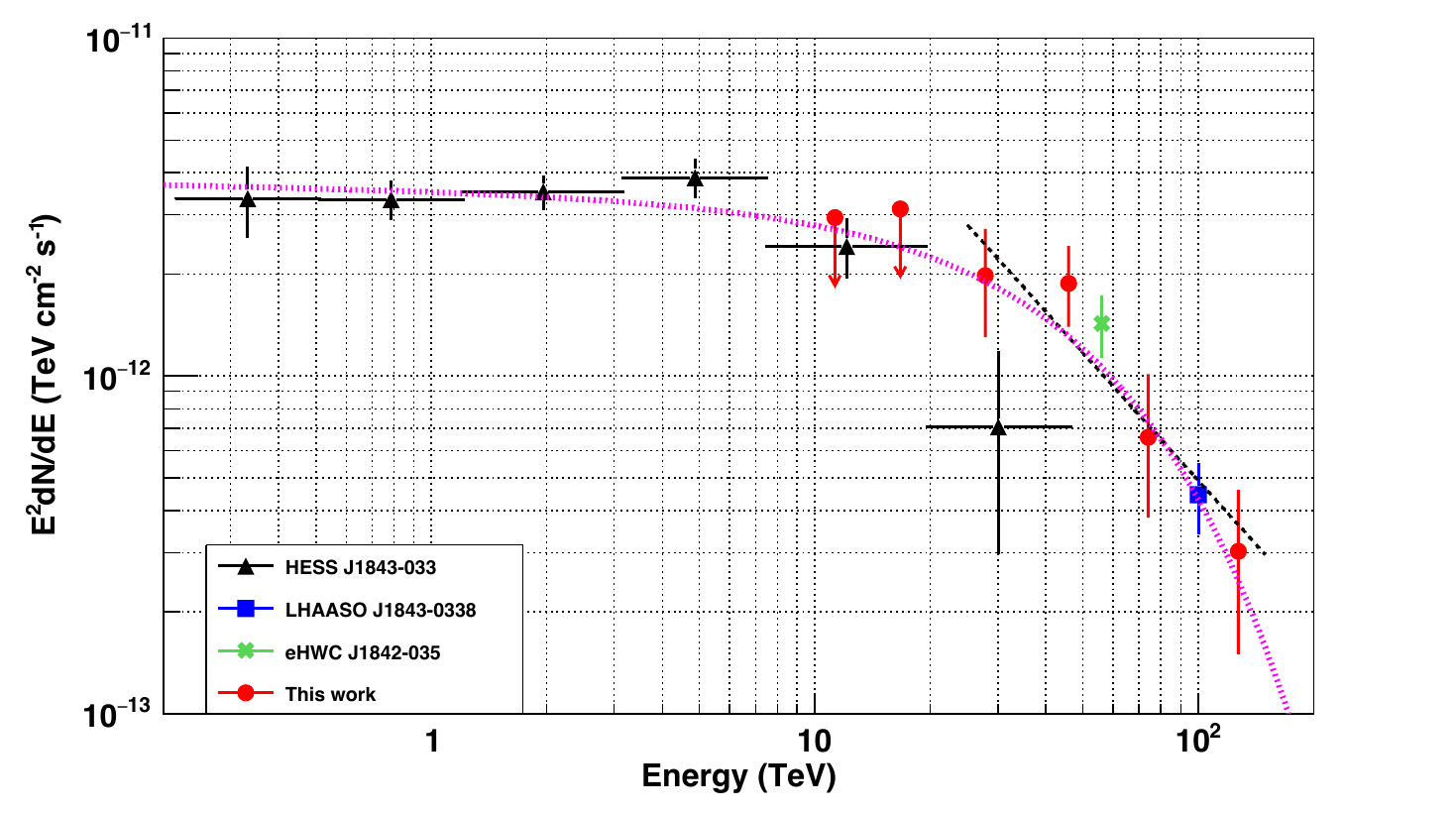}
  \caption{Differential gamma-ray energy spectrum measured in this work (red). Vertical bars and downward arrows denote $1\, \sigma$ statistical errors and the $95\%$ upper limits, respectively. The black dashed line shows the best-fit power-law function to our results in $25\, {\rm TeV}<E<130\, {\rm TeV}$. Also shown are the flux points of HESS J1843-033 (black, \citet{HGPS2018}), eHWC J1842-035 (light green, \citet{HAWC56TeV}), and LHAASO J1843-0338 (blue, \citet{LHAASO100TeV}). The flux of eHWC J1842-035 is calculated from the integral flux above $56\, {\rm TeV}$ assuming a differential spectral index of $-2.7$. The magenta dotted curve shows the best-fit power-law function with an exponential cutoff fitted to the combined spectra of HESS J1843-033, LHAASO J1843-0338, and TASG J1844-038.}
  \label{enesp}
\end{figure}

Figure \ref{enesp} shows the gamma-ray energy spectrum measured in this work. A gamma-ray flux is calculated only if the statistical significance of gamma-ray detection in each energy bin exceeds $2\, \sigma$, and otherwise, a $95\%$ upper limit is calculated. Our results in $25\, {\rm TeV}<E<130\, {\rm TeV}$ can be fitted with a power-law function of ${\rm d}N/{\rm d}E = (9.70\pm 1.89)\times 10^{-16} (E/40\, {\rm TeV})^{-3.26\pm 0.30}\, {\rm TeV}^{-1} {\rm cm}^{-2} {\rm s}^{-1}$ with $\chi^2/{\rm d.o.f.} = 2.1/2$. As a possible systematic uncertainty, the uncertainty in the source extension $\sigma_{\rm ext}$ affects the flux normalization by $19\%$ and the spectral index by $0.02$. The absolute energy-scale uncertainty of $12\%$ \citep{Amenomori_et_al_2009} also affects the flux normalization by $39\%$. Summing up these two in quadrature, the total systematic uncertainty in the flux normalization is calculated as $43\%$. The spectrum of TASG J1844-038 smoothly connects to that of HESS J1843-033 and is consistent with the fluxes of eHWC J1842-035 and LHAASO J1843-0338, supporting the common origin for gamma rays from these sources. A fit of a simple power-law function to the combined spectra of HESS J1843-033, LHAASO J1843-0338, and TASG J1844-038 is disfavored at the $5.0\, \sigma$ level, and the spectrum is better described with the following power-law function with an exponential cutoff:
\begin{equation}
  \frac{{\rm d}N}{{\rm d}E} = N_0 \bigg(\frac{E}{{\rm TeV}}\bigg)^{-\Gamma} {\rm exp}\bigg(-\frac{E}{E_{\rm cut}}\bigg)
\end{equation}
where the best-fit parameters are $N_0 = (3.57\pm 0.26) \times 10^{-12}\, {\rm TeV}^{-1} {\rm cm}^{-2} {\rm s}^{-1}$, $\Gamma = 2.02\pm 0.06$ and $E_{\rm cut} = 49.5\pm 9.0 \, {\rm TeV}$ with $\chi^2/{\rm d.o.f.} = 10.4/8$. The flux of eHWC J1842-035 is not included in the combined fit because only the integral flux above 56 TeV is presented by \citet{HAWC56TeV} (the differential flux of eHWC J1842-035 shown in Figure \ref{enesp} is calculated from the integral flux with a spectral index of $-2.7$ assumed in their paper).

The time variation of the gamma-ray flux is also searched by equally dividing the data into four independent periods and calculating the integral gamma-ray flux in each period. A constant function is fitted to the gamma-ray flux points over the four periods, and the best-fit value is $(4.65 \pm 0.89)\times 10^{-14}\, {\rm cm}^{-2} {\rm s}^{-1}$ above $25\, {\rm TeV}$ with $\chi^2/{\rm d.o.f.} = 1.9/3$, consistent with the null hypothesis that no time variation of the gamma-ray flux takes place during our observation. A similar result is obtained for the flux above $40\, {\rm TeV}$: $(1.59 \pm 0.38)\times 10^{-14}\, {\rm cm}^{-2} {\rm s}^{-1}$ with $\chi^2/{\rm d.o.f.} = 3.2/3$.

\clearpage
\section{Discussion} \label{sec:dis}
\begin{figure}[b]
  \centering
  \includegraphics[scale=0.8]{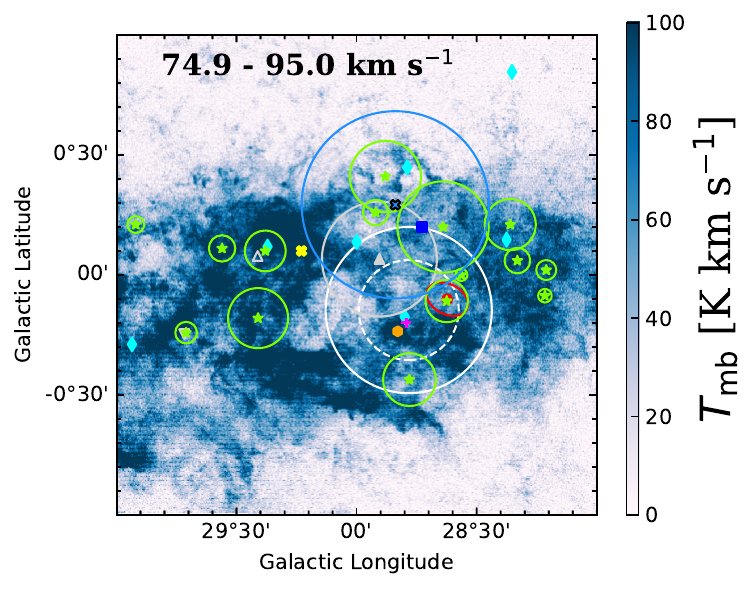}
  \caption{Integrated intensity map of $^{12}{\rm CO}$ ($J=1-0$) line emission in the velocity range from $75\, {\rm km}\, {\rm s}^{-1}$ to $95\, {\rm km}\, {\rm s}^{-1}$ reconstructed from the public FUGIN data \citep{FUGIN}. The color scale shows the main beam brightness temperature. The solid and dashed white circles denote the extension of TASG J1844-038 with a radius of $\sigma_{\rm ext}$ and the positional uncertainty at the $68\%$ confidence level with a radius of $R_{0.68}$, respectively (see Table \ref{tab:posinfo}). Positions and extensions of nearby sources are shown in the same way as in Figure \ref{sigmap1} except that the orange hexagon denotes the star-forming region N49 \citep{Dirienzo_et_al_2012}.}
  \label{12COmap}
\end{figure}

Among the sources near TASG J1844-038, SNR G28.6-0.1, whose center is within the positional error circle of TASG J1844-038 as shown in Figure \ref{12COmap}, is one of the prominent candidates for the counterpart. A radio complex G28.60-0.13 was first discovered by \citet{Helfand_et_al_1989}, some regions of which (regions C and F) were found to be emitting non-thermal radio emissions. Subsequently, an X-ray counterpart AX J1843.8-0352 was detected by ASCA and Chandra \citep{Bamba_et_al_2001, Ueno_et_al_2003}. The authors claimed that the observed non-thermal radio and X rays are synchrotron radiation of high-energy electrons accelerated by a shell-type SNR. The extension of AX J1843.8-0352 (mean diameter of $9'$ \citep{Ueno_et_al_2003}) is smaller than that of TASG J1844-038 at the $2.3\, \sigma$ level. The distribution of gamma rays of leptonic origin is expected to have approximately the same extension as that of the observed non-thermal X rays. The possible discrepancy thus suggests that, if SNR G28.6-0.1 is the counterpart of TASG J1844-038, radiation from hadronic cosmic rays accelerated by the SNR contributes to the observed gamma-ray emission. The maximum energy of accelerated cosmic-ray protons could be $\simeq 500\, {\rm TeV}$, roughly an order of magnitude larger than the cutoff energy of $\simeq 50\, {\rm TeV}$ in the gamma-ray energy spectrum. In addition, the SNR is found to be interacting with molecular clouds in the velocity channel of around $86\, {\rm km}\,{\rm s^{-1}}$ \citep{Ranasinghe_and_Leahy_2018}. Considering the peculiar motion of the Galactic objects \citep{Reid_2014}, the distribution of molecular clouds in the integrated velocity channel from $75\, {\rm km}\, {\rm s}^{-1}$ to $95\, {\rm km}\, {\rm s}^{-1}$ is investigated with $^{12}{\rm CO}$ ($J=1-0$) line emission using the public FUGIN data \citep{FUGIN}. The result is shown in Figure \ref{12COmap} and some clouds are found to overlap with the TASG J1844-038 region. From the aforementioned reasons and the estimated age of SNR G28.6-0.1 ($2.7\, {\rm kyr}$ by \citet{Bamba_et_al_2001} or $19\, {\rm kyr}$ by \citet{Ranasinghe_and_Leahy_2018}), this SNR could be a similar system to a PeVatron candidate SNR G106.3+2.7 \citep{TibetSNRG106} and thus could have been accelerating protons up to the PeV energy range in the past. Here the diffusion time scale of cosmic-ray protons in molecular clouds is estimated following \citet{Gabici_et_al_2007_2}:
\begin{equation}\nonumber
  \tau_{\rm diff} \sim 1.2\times 10^{4}\, \chi^{-1} \bigg(\frac{R_{\rm sys}}{20\, {\rm pc}}\bigg)^2 \bigg(\frac{E}{\rm GeV}\bigg)^{-0.5} \bigg(\frac{B}{10\, \mu{\rm G}}\bigg)^{0.5}\, {\rm yr},
\end{equation}
where $R_{\rm sys}$ is the radius of the system, $B$ the magnetic field strength, and $\chi$ the factor indicating the suppression of diffusion from the Galactic average. The extension of TASG J1844-038 corresponds to $\simeq 60\, {\rm pc}$ at the distance of $9.6\, {\rm kpc}$ which is an estimated distance to SNR G28.6-0.1 \citep{Ranasinghe_and_Leahy_2018}, and the magnetic field can be assumed as $10\, \mu{\rm G}$ \citep{Crutcher_1991}. Cosmic-ray protons with $E\gtrsim 250\, {\rm TeV}$, which will generate $\pi^{0}$-decay gamma rays with $E\gtrsim 25\, {\rm TeV}$ in interactions with the clouds, can diffusively propagate up to the size of TASG J1844-038 within $\tau_{\rm diff} \sim 2\, {\rm kyr}$ even if the case of $\chi \sim 0.1$ is considered. Under the same condition, the extension of HESS J1843-033 can be similarly explained by the gamma-ray emission from the accelerated protons with $E \simeq 10\, {\rm TeV}$ diffusing for $\tau_{\rm diff} \sim 5\, {\rm kyr}$. These diffusion time scales do not conflict with the estimated age of SNR G28.6-0.1.

Another prominent candidate is the pulsar PSR J1844-0346 nearly centered at TASG J1844-038. This pulsar, also named 4FGL J1844.4-0345 \citep{4FGLCatalog}, was discovered by {\it Fermi}-LAT \citep{C_J_Clerk_et_al_2017}, and its spin-down luminosity, characteristic age, and spin period are $4.2\times10^{36} \, {\rm erg\, s^{-1}}$, $12\, {\rm kyr}$, and $113\, {\rm ms}$, respectively. Assuming the distance to the pulsar as $4.3\, {\rm kpc}$ \citep{Devin_et_al_2021}, the gamma-ray energy flux of $1.1\times 10^{-11}\, {\rm erg}\, {\rm cm}^{-2}\, {\rm s}^{-1}$ in $1\, {\rm TeV}<E<10\, {\rm TeV}$ \citep{HGPS2018} is translated into the luminosity of $2.4\times 10^{34}\, {\rm erg}\, {\rm s}^{-1}$. The aforementioned properties of PSR J1844-0346 and the surrounding gamma-ray emission including its flat spectral index of $2.02\pm 0.06$ in the TeV range are similar to those of ``TeV pulsar wind nebulae (PWNe)'' \citep{HESS_TeVPWN_2017} as proposed by \citet{Sudoh_et_al_2021}. The energies of electrons that generate gamma rays with $E \simeq 50\, {\rm TeV}$ by Inverse Compton Scattering (ICS) off the cosmic microwave background (CMB) photons are $\simeq 90\, {\rm TeV}$ (see the equation (10) of \citet{Hinton_Hofmann_2009}). Assuming the magnetic field as $3\, \mu{\rm G}$ and the diffusion coefficient of the $90\, {\rm TeV}$ electrons as $4.4\times 10^{27}\, {\rm cm}^2\, {\rm s}^{-1}$ following the observation of Geminga \citep{HAWC_Geminga}, the time scale of the electrons diffusing up to the size of TASG J1844-038 ($\simeq 26\, {\rm pc}$ at the assumed distance to PSR J1844-0346 of $4.3\, {\rm kpc}$) is estimated at $\sim 8\, {\rm kyr}$, which is within the characteristic age of PSR J1844-0346 and the cooling time of $\sim 11\, {\rm kyr}$ due to the synchrotron emission and ICS estimated from the equations (5) and (7) of \citet{Hinton_Hofmann_2009}. Similarly, the extension of HESS J1843-033 in the TeV range can be explained by ICS off the CMB photons by $\simeq 10\, {\rm TeV}$ electrons diffusing for $\sim 8\, {\rm kyr}$. \citet{Devin_et_al_2021} found no radio nor X-ray emission that indicates the existence of a PWN. Given the characteristic age of PSR J1844-0346, synchrotron emission from the PWN would not be bright enough to be observed due to the decay of the magnetic field \citep{Tanaka_Takahara_2010}. Future studies with wide field-of-view and high-sensitivity observations of X rays will be a key to constraining the TeV PWN scenario.

It should be noted that there are additional SNR candidates near TASG J1844-038 discovered by THOR \citep{Anderson_et_al_2017}. Out of these candidates, G28.56+0.00, G28.64+0.20, and G28.78-0.44 overlap with TASG J1844-038, and future research on these SNRs is expected. Moreover, as pointed out by \citet{Devin_et_al_2021}, the star-forming region N49 \citep{Dirienzo_et_al_2012} is also located within TASG J1844-038 (see Figure \ref{12COmap}). Several observations support the acceleration of cosmic rays in star-forming regions (for example, see \citet{Ackermann_et_al_2011, Aharonian_et_al_2019}), and detailed morphological studies of the gamma-ray emission of TASG J1844-038 by imaging atmospheric Cherenkov telescopes and its comparison with the stellar and gas distributions observed in other wavelength ranges will be needed to discuss a possible association with N49.

\section{Conclusion} \label{sec:con}
A gamma-ray source TASG J1844-038 is detected above $25\, {\rm TeV}$ near HESS J1843-033 with a statistical significance of $6.2\, \sigma$ using the data recorded by the Tibet AS array and the MD array. Its extension is estimated at $0{\fdg}34\pm 0{\fdg}12$, and the center $({\alpha}, \delta) = (281{\fdg}09\pm 0{\fdg}10, -3{\fdg}76\pm 0{\fdg}09)$ is statistically consistent with those of HESS J1843-033, eHWC J1842-035, and LHAASO J1843-0338. The gamma-ray energy spectrum is measured beyond $100\, {\rm TeV}$ for the first time and is found to be smoothly connected to that of HESS J1843-033. The combined spectra of HESS J1843-033, LHAASO J1843-0338, and TASG J1844-038 are well fitted with a power-law function with the exponential cutoff energy of $49.5\pm 9.0\, {\rm TeV}$.

The origin of TASG J1844-038 is also discussed in detail for the first time assuming its associations with SNR G28.6-0.1 and PSR J1844-0346. If SNR G28.6-0.1 is assumed to be the counterpart, the nature of TASG J1844-038 can be explained by $\pi^{0}$-decay gamma rays generated in hadronic interactions between adjacent molecular clouds and cosmic-ray protons with $E \lesssim 500\, {\rm TeV}$ that are accelerated by the SNR and diffusely propagate through the clouds. Given the similarities with SNR G106.3+2.7 in terms of the maximum energy of accelerated protons, the partial overlap of the gamma-ray distribution with molecular clouds, and the SNR's age, SNR G28.6-0.1 could have been a PeVatron and accelerating cosmic-ray protons up to the PeV energy range in the past. On the other hand, if associated with PSR J1844-0346, TASG J1844-038 can be explained by the gamma-ray emission from a TeV PWN generated by ICS off the CMB photons by high-energy electrons. Other nearby sources including SNR candidates and the star-forming region N49 should also be studied to investigate their contribution to the observed gamma-ray emission.

\begin{acknowledgments}
  The collaborative experiment of the Tibet Air Shower Arrays has been conducted under the auspices of the Ministry of Science and Technology of China and the Ministry of Foreign Affairs of Japan. This work was supported in part by a Grant-in-Aid for Scientific Research on Priority Areas from the Ministry of Education, Culture, Sports, Science and Technology, and by Grants-in-Aid for Science Research from the Japan Society for the Promotion of Science in Japan. This work is supported by the National Key R\&D Program of China (No. 2016YFE0125500), the Grants from the National Natural Science Foundation of China (No. 11533007, No. 11673041, No. 11873065, No. 11773019, No. 11773014, No. 11633007, No. 11803011, and No. 11851305), and the Key Laboratory of Particle Astrophysics, Institute of High Energy Physics, CAS. This work is also supported by the joint research program of the Institute for Cosmic Ray Research (ICRR), the University of Tokyo. S. Kato is supported by JST SPRING, Grant Number JPMJSP2108. This publication makes use of data from FUGIN, FOREST Unbiased Galactic plane Imaging survey with the Nobeyama 45-m telescope, a legacy project in the Nobeyama 45-m radio telescope.
\end{acknowledgments}

\appendix

\section{Monte Carlo simulation and optimization of muon-cut event selection} \label{app:MC}
A total number of $1.1\times 10^8$ gamma rays are generated in $300\,{\rm GeV}<E<10\,{\rm PeV}$ using Corsika v7.4000 \citep{CORSIKA} following a simple power-law spectrum with an index of $-2.0$ along the path in the sky of a hypothetical source with $\delta = 0^{\circ}$. These gamma rays are injected into the atmosphere, and the development of the extensive air showers is simulated until they reach the site of the experiment. Shower cores are randomly thrown over the circular region with a radius of $300\, {\rm m}$ centered at the Tibet AS array. The detector responses to the showers are simulated with Geant4 v10.0 \citep{GEANT4}. The detector simulation includes the calculation of energy loss of shower particles in the plastic scintillation detectors, the soil, and the concrete layers of the MD array, and their Cherenkov-light emission in the water layer of the MD array. The output is fed into our analysis pipeline in the same way as experimental data. In the Monte Carlo data analysis, the gamma-ray events are appropriately weighted accounting for the spectral index of the source. The resultant angular and energy resolutions for $100\, \rm TeV$ gamma rays from HESS J1843-033 are estimated at $\simeq 0{\fdg}28$ ($50\%$ containment) and $\simeq 30\%$, respectively. Compared with the previous work \citep{TibetCrab}, these resolutions worsen by $\simeq 40\%$ due to the large meridian zenith angle of HESS J1843-033 ($\simeq 33^{\circ}$) and the extension of the analyzed zenith-angle range from $40^{\circ}$ to $50^{\circ}$ (see Section \ref{sec:experiment}).

The muon cut is optimized for this work using the Monte Carlo gamma-ray data and experimental background data. Following the equi-zenith angle method \citep{Amenomori_et_al_2003}, the numbers of background events are averaged over 100 OFF regions opened around the ON-source region which contains the HESS J1843-033 region. Both the ON-source and OFF regions are circular regions with a radius of $0{\fdg}7$. The number of gamma-ray events is normalized to that of excess counts in the ON-source region. The gamma-ray and background events are then binned in $\Sigma \rho$ which denotes the total number density of shower particles recorded with the plastic scintillation detectors of the AS array, and the optimum cut on the number of muons ($\Sigma N_{\mu}$) is determined so that the figure of merit $S/\sqrt{S+B}$ ($S$ is the number of gamma-ray events and $B$ that of background events) is maximized in each bin. The resultant muon cut requires events to satisfy $\Sigma N_{\mu} < 1.8\times 10^{-3} (\Sigma \rho/{\rm m}^{-2})^{1.1}$ or $\Sigma N_{\mu} < 0.4$. After applying the muon cut to events, the survival ratio of gamma-ray events and the rejection power of background events are estimated at $\simeq 80\%$ and $\gtrsim 99.9\%$, respectively, at the gamma-ray equivalent energy of $100\, {\rm TeV}$. Due to the tight muon cut to improve the figure of merit for the low gamma-ray flux, the gamma-ray survival ratio worsens by $\simeq 10\%$ from the previous work \citep{TibetCrab}.

\section{Calculation of positional uncertainties in TASG J1844-038 and nearby sources}\label{app:calcdev}
Uncertainties in the positions of TASG J1844-038, eHWC J1842-035, and LHAASO J1843-0338 are evaluated in terms of the error radius $R_{0.68}$ at the $68\%$ confidence level defined similarly as that in \citet{HGPS2018} but employing the equatorial coordinates $(\alpha, \, \delta)$ instead of the Galactic coordinates $(l, \, b)$:
\begin{equation}\label{eq:R}
  R_{0.68} = f_{0.68} \sqrt{\Delta{\alpha}^2_{\rm stat} + \Delta{\alpha}^2_{\rm sys} + \Delta\delta^2_{\rm stat} + \Delta\delta^2_{\rm sys}}
\end{equation}
where $f_{0.68} = \sqrt{-2\, {\rm ln}(1-0.68)}$ \citep{Abdo_et_al_2009}, and $\Delta\alpha_{\rm stat (\rm sys)}$ and $\Delta\delta_{\rm stat (\rm sys)}$ are the statistical (systematic) uncertainties in the right ascension and declination, respectively. For HESS J1843-033, HESS J1844-030, and HESS J1846-029, their $R_{0.68}$'s are cited from \citet{HGPS2018}.

%\bibliography{mybibfile}{}
%\bibliographystyle{aasjournal}

\end{document}